# DAN-Net: Dual-Domain Adaptive-Scaling Non-local Network for CT Metal Artifact Reduction

Tao wang[1], Wenjun Xia[1], Yongqiang Huang[1], Huaiqiang Sun[2], Yan Liu[3], Hu Chen[1], Jiliu Zhou[1], Yi Zhang[1]*

1. College of Computer Science, Sichuan University, Chengdu 610065, China
2. Department of Radiology, West China Hospital of Sichuan University, Chengdu 610041, China
3. College of Electrical Engineering, Sichuan University, Chengdu 610065, China

**Abstract**: Metal implants can heavily attenuate X-rays in computed tomography (CT) scans, leading to severe artifacts in reconstructed images, which significantly jeopardize image quality and negatively impact subsequent diagnoses and treatment planning. With the rapid development of deep learning in the field of medical imaging, several network models have been proposed for metal artifact reduction (MAR) in CT. Despite the encouraging results achieved by these methods, there is still much room to further improve performance. In this paper, a novel Dual-domain Adaptive-scaling Non-local network (DAN-Net) for MAR. We correct the corrupted sinogram using adaptive scaling first to preserve more tissue and bone details as a more informative input. Then, an end-to-end dual-domain network is adopted to successively process the sinogram and its corresponding reconstructed image generated by the analytical reconstruction layer. In addition, to better suppress the existing artifacts and restrain the potential secondary artifacts caused by inaccurate results of the sinogram-domain network, a novel residual sinogram learning strategy and nonlocal module are leveraged in the proposed network model. In the experiments, the proposed DAN-Net demonstrates performance competitive with several state-of-the-art MAR methods in both qualitative and quantitative aspects.

---

* Corresponding author.
  E-mail address: yzhang@scu.edu.cn.

# 1. Introduction

Computed tomography (CT) technology has developed rapidly in clinical, industrial, security and other spheres [1]. With the help of CT images, medical diagnosis and treatments can be conducted effectively. However, the effects of noise, photon starvation, beam hardening, scattered radiation and nonlinear partial volume effects are much more severe in the case of metallic implants in scanned regions [2]. Due to these metallic objects, the reconstructed CT images are contaminated by heavy artifacts, specifically those called "metal artifacts." These artifacts degrade the imaging quality and severely compromise doctors' diagnoses. In particular, some artifacts and certain lesions have considerable commonalities, leading to misdiagnosis, and subsequent medical image analysis is difficult [3]. Therefore, it is of great significance to reduce metal artifacts in CT images.

During the past several decades, numerous metal artifact reduction (MAR) methods have been dedicated to addressing the abovementioned problem. Conventional MAR methods can be grouped into three categories: projection completion methods, iterative reconstruction methods and image postprocessing methods [4]. The projection completion methods regard projection data in the metal trace as missing information and fill in lost data with estimated values by different image inpainting [5-7] or interpolation strategies [8-12]. Linear or polynomial interpolations [8, 9, 12] are widely adopted to estimate the missed values in projection data. However, interpolation-based methods can hardly guarantee smoothness at the interpolation boundaries [13]. After filtering, discontinuities are amplified at the metal trace boundaries, which introduce new artifacts into the reconstructed CT images. To fully explore the local information in both dimensions of the angle and detector bin, some diffusion-based image inpainting methods were introduced for projection completion [5-7]. Although these methods may mitigate the discontinuity to some extent, extra artifacts are still inevitable in the reconstructed images.

To smooth the transition region between the metal and nonmetal portions and to suppress secondary artifacts, some prior image-based methods have been proposed [14-16], such as the normalized metal artifact reduction (NMAR) method [15]. NMAR normalizes the projection data with the constraint of prior images obtained by multi-threshold segmentation based on interpolation methods. Corrected CT images can be derived from completed sinograms by filtered back projection (FBP). However, the result of NMAR is limited by the quality of the prior image. In addition, FBP is based on the line integral model, which does not take into account the statistical characteristics of measured data and simply assumes that the measured data are noiseless and that all response lines have the same weight, which is not always consistent with the real situation. Iterative reconstruction is an alternative way to tackle these problems, which improves image quality gradually based on constrained optimization, such as the least square method and maximum likelihood. Classical iterative reconstruction MAR methods can be divided into two groups. One uses projection data outside of the metal trace, which can be regarded as clean data [11, 17-22]. The other adopts a statistical objective function to decay corrupted projection data [23]. However, iterative methods are usually time-consuming and require manually well-designed regularizers [24, 25], both of which bring difficulties to clinical application. Image postprocessing methods [26, 27] aim to reduce metal artifacts in the image domain without accessing raw projection data. However, since the noise and artifacts in CT images do not obey any specific statistical distribution, postprocessing methods usually cannot suppress the artifacts well and are apt to distort the anatomic structure [4, 28].

Recently, with the successful applications of deep learning (DL) in many fields [29-33], DL-based methods have shown great potential for medical imaging [34, 35]. Recently, several DL-based MAR methods have been proposed. Different network architectures, such as convolutional neural networks (CNNs) and generative adversarial networks (GANs), have been utilized to recover the missed data in the metal trace [36-40]. Meanwhile, some studies have been dedicated to using DL methods to reduce metal artifacts in the image domain. Zhang et al. [41] proposed a CNN framework (CNNMAR) fusing different MAR methods to improve the performance of artifact reduction. To eliminate metal artifacts from original CT images, [42] introduced a novel unsupervised artifact disentanglement network (ADN). Gjesteby et al. [43] took detailed images derived from filtering and base images by NMAR as inputs and mapped them to metal artifact-free images with a dual-stream residual network.

Despite the encouraging results achieved by the abovementioned sinogram- or image domain-based DL methods, there are still some limitations in single-domain methods, which has been demonstrated in [44]. For sinogram domain-based DL methods, although corrupted projections within metal traces are local, it is difficult to preserve continuity at metal trace boundaries, where secondary artifacts can be introduced easily. In terms of the image domain-based methods, the input CT images reconstructed from corrupted sinograms are full of severe artifacts, which cover most clinically important details. These images with low quality may lead to misclassification of some structures and artifacts due to their similar patterns. In addition, artifacts are nonlocal in the image, which is hard to remove completely. Thus, the goal for MAR becomes twofold. The first goal is to eliminate existing artifacts as much as possible, and the other is to avoid introducing extra artifacts. To this end, combining the merits of both projection and image domain-based methods is meaningful. Some end-to-end dual domain networks were proposed very recently. Lin et al. [44] proposed DuDoNet, which progressively restores sinogram consistency and enhances CT images linked by a differentiable radon inversion layer. Lyu et al. [45] proposed improving DuDoNet by specifying the metal mask projection and encoding it into the network. Yu et al. [28] proposed employing an image-domain network to generate a prior image at first. Then, the sinogram obtained from the prior image was utilized to guide the sinogram-domain network. In [46], the authors proposed using partial convolution to recover irregular metal trace regions with only valid pixels outside the corrupted areas. Furthermore, an auxiliary inpainting network is introduced to suppress the secondary artifacts in the reconstructed image from the previous step. Both sinograms from the last two steps were fused to generate the final result.

Due to their state-of-the-art performance, dual-domain networks have become the mainstream for MAR. However, current dual domain-based methods still suffer from some critical limitations. [44, 46] regarded projection data in the metal trace as missing data, which resulted in the loss of details near the metal area in reconstructed CT images. [45] and [28] used metal corrupted projection data and corresponding reconstructed CT images as inputs directly. Actually, the data in the metal trace have a much higher amplitude than the data outside the metal trace, and there is a rapid change at the boundary of the metal trace. According to the CT imaging principle [45], due to this amplitude difference, data inside and outside of the metal trace can be regarded as obeying two different data distributions. It is difficult for neural networks to transform two different data distributions into a uniform distribution. As demonstrated in [44], the authors experimentally found that their method did not perform well while taking original sinogram and CT images as inputs. Meanwhile, the change in the boundary will cause weak continuity of the first derivative of projection data in a certain section, which will be further expanded by filtering and will generate extra artifacts [47].

To address the problems mentioned above, in this paper, a novel Dual-domain Adaptive-scaling Non-local network (DAN-Net) for MAR is proposed. The projection data are considered to be composed of two parts: one part comes from the tissues and the other part comes from the metal objects. A rough estimation of tissue-like projection data in the metal trace is obtained by a linear interpolation operation, and the residual between it and the original projection data is regarded as the contribution of the metal.

To weaken the rapid change caused by metal implants and retain the data characteristics of this part, the residual in the metal trace is adaptively scaled [48, 49] and then filtered by an average filter to further improve the continuity at metal trace boundaries. The results of this adaptive scaling and corresponding reconstruction by FBP are used as the inputs of our network. In addition, a novel residual sinogram learning strategy is applied in the sinogram-domain network to weaken the rapid change in projection data and improve the smoothness of the projection. On the other hand, to handle the nonlocality of artifacts, a nonlocal U-Net architecture is employed for image-domain enhancement, capturing long-range dependencies via nonlocal operations. The whole network is trained in an end-to-end manner so that the image-domain enhancement and sinogram-domain enhancement can benefit each other.

Our main contributions are summarized as follows.

1) Different from current dual-domain networks, the original sinogram is preprocessed using adaptive scaling and accompanied by its corresponding FBP result as the inputs, which can preliminarily suppress metal artifacts.

2) A novel residual sinogram learning strategy is proposed to improve the smoothness of the projection and alleviate the secondary artifacts in the reconstructed CT images.

3) A nonlocal U-Net architecture is designed for image-domain enhancement, which can capture long-range dependencies of metal artifacts and further improve image quality.

The remainder of this paper is organized as follows. The proposed DAN-Net is elaborated in Section 2. The experiments on and results for the simulated and clinical data are presented in Section 3. The results of analytical studies are shown in Section 4. Discussion and the conclusion are provided in Section 5.

**2. Method**

## 2.1 Problem formulation

In our work, we consider the case of a 2D attenuation distribution. If there are metallic objects in the scanner field, the linear attenuation coefficients can be expressed as follows:

$$\mu(E) = \mu_t(E) \odot (1 - M) + \mu_m(E) \odot M \tag{1}$$

where $\mu_t(E)$ and $\mu_m(E)$ represent attenuation images of normal human tissue and metal parts, respectively; $M$ denotes the metal mask in the CT image; and $\odot$ is the elementwise multiplication. In this case, letting $\mathcal{P}$ be a forward projection operation, the projection data contaminated by metals, $S_{ma}$, can be calculated as follows [45]:

$$\begin{aligned} S_{ma} &= -\ln \int \eta(E) \exp\left(-\mathcal{P}(\mu(E))\right) \, dE \\ &= -\ln \int \eta(E) \exp\left(-\mathcal{P}((\mu_t(E) \odot (1-M) + \mu_m(E) \odot M))\right) \, dE \\ &= -\ln \int \eta(E) \exp\left(-\mathcal{P}(\mu_t(E) \odot (1-M))\right) \, dE - \ln \int \eta(E) \exp\left(-\mathcal{P}(\mu_m(E) \odot M)\right) \, dE \\ &= S_{tissue} + S_{metal} \end{aligned} \tag{2}$$

where $\eta(E)$ denotes the intensity distribution with spectral energy at $E$. Thus, $S_{ma}$ can be regarded as having two parts: one is contributed by the attenuation of tissues, denoted as $S_{tissue}$, and the other is produced by metal objects, denoted as $S_{metal}$. The tissue-like projection data in the metal trace, referred to as $S_{LI}$, are obtained by performing linear interpolation, and the residual of the original projection data and linear interpolation results, notated as $S_{sub} = S_{ma} - S_{LI}$, are regarded as the metal contribution. In most MAR methods, the target is to remove $S_{metal}$ from $S_{ma}$. Ideally, we assume $S_{LI} = S_{tissue}$ and $S_{sub} = S_{metal}$. However, $S_{LI}$ is just a coarse estimation of $S_{tissue}$, and some useful information is still reserved in $S_{sub}$. Moreover, as shown in Eq. (1), if we simply discard the projection data in the metal trace, both tissue and metal projections will be lost, and the reconstructed CT image has to take the risk of losing tissue details around the metallic implants. Based on these considerations, our method attempts to restore this valuable information $S_v$ from $S_{sub}$. Then, the artifact-suppressed image $X_{sino}$ can be reconstructed from $S_{LI} + S_v$, which can be expressed as $X_{sino} = \mathcal{P}^{-1}(S_{LI} + S_v)$, in which $\mathcal{P}^{-1}$ denotes FBP operation.

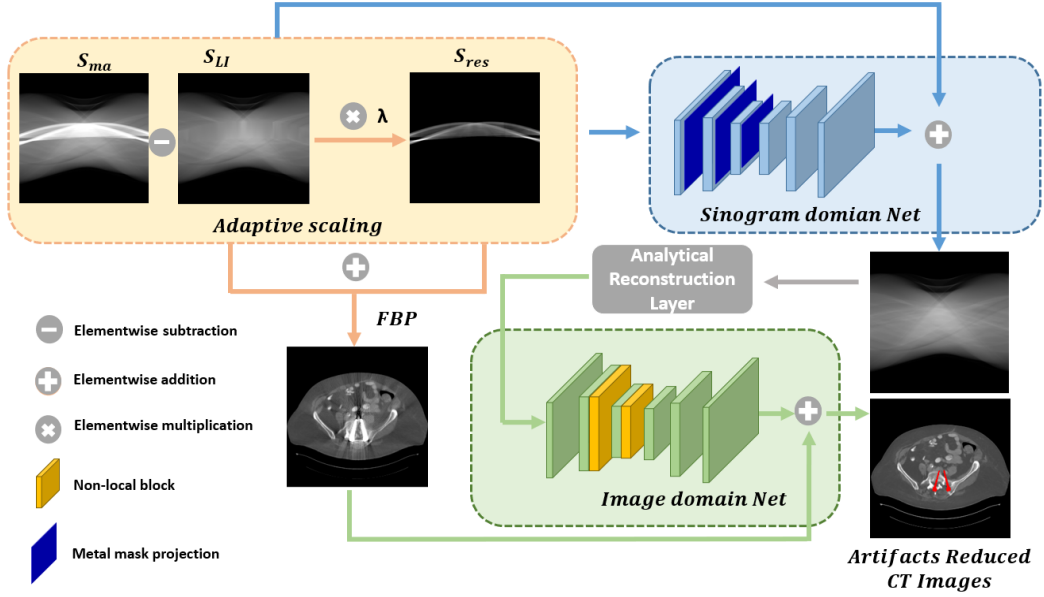

Fig. 1: Overview of the proposed DAN-Net.

**2.2 The proposed DAN-Net**

To simultaneously leverage the advantages of both sinogram- and image-domain information, we adopt a dual-domain joint learning strategy for CT MAR, and back-propagation of gradients is conducted by the analytical reconstruction layer. Fig. 1 depicts the overview of our proposed DAN-Net, which consists of three components: adaptive scaling, sinogram-domain network and image-domain network. More details are presented in subsequent sections.

**2.2.1 Adaptive scaling**

When X-rays pass through a metal material with high attenuation coefficients, the intensity of low energy will be significantly reduced. At this time, the beam-hardening effect will be more pronounced, leading to an abrupt change in projection data at metal trace boundaries. As we mentioned before, this change will raise more artifacts after filtering. To eliminate the rapid shift in projection data caused by the metal and maintain more useful information, Chen et al.[50] adopted a linear attenuation operation to restore the data in the metal trace, which can be written as the following formula for simplicity:

$$S_{res} = \lambda * S_{sub} \tag{3}$$

$$S_{pre} = S_{LI} + S_{res} \tag{4}$$

where $\lambda$ is the scaling parameter to control the trade-off between artifact reduction and detail preservation around the metallic implant in the final reconstructed CT images. $S_{res}$ and $S_{pre}$ represent the scaled metal projection and the corrected projection after adaptive scaling, respectively. As a result, a greater $\lambda$ will keep more tissue details but lead to more artifacts as well, while a smaller $\lambda$ will generate fewer artifacts but lose more tissue details. Typically, the value of $\lambda$ is set to between 0.3 and 0.5 according to [50]. We chose $\lambda = 0.4$ experimentally and corresponding adaptively scaled CT images obtained as $X_{pre} = \mathcal{P}^{-1}(S_{pre})$. Fig. 2 shows one example of the original sinogram, an LI corrected sinogram and our adaptively scaled sinogram, as well as their corresponding reconstructions. Although the adaptively scaled result (D2) has more artifacts than the LI corrected output (C2), many more bone and tissue details, especially bounding the metal, are reserved. It will be helpful to compensate for the inaccurate beam hardening corrections and reduce the impact of any errors within the projections of metal implants when backprojecting them to other bone or tissue positions in the reconstructed image.

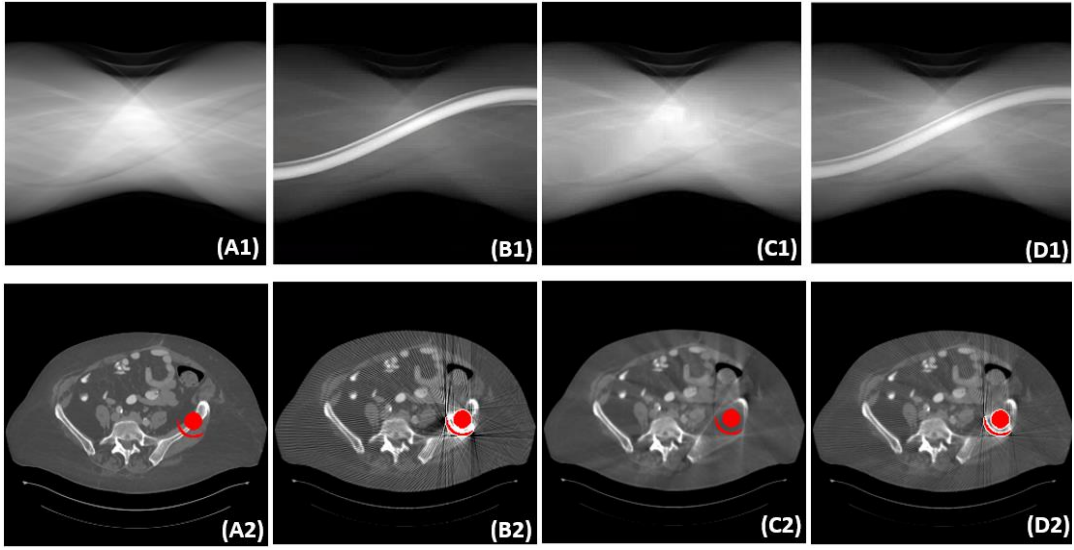

Fig. 2: Visual comparison with LI and adaptive scaling. (A1) and (A2) refer to the sinogram image and the referenced CT image. (B1) and (B2) are the corrupted sinogram and reconstruction. The MAR sinogram and CT image using LI (C1) and (C2) and adaptive scaling (D1) and (D2) are also presented. The simulated metal masks are colored in red for better visualization.

**2.2.2 Sinogram Domain Network**

To complete the sinogram, we train a neural network to process the projection data, denoted as $G_{sino}$. If we take only the LI corrected sinogram $S_{LI}$ as the input of $G_{sino}$, due to the characteristics of $S_{LI}$, the CT image reconstructed from the output of $G_{sino}$ will be oversmoothed, and some tissue details will be lost [45]. On the other hand, it is challenging to restore information directly from the original corrupted sinogram because the projection data inside and outside the metal trace follow two different distributions. Moreover, as we depicted before, there is some useful information in the adaptively scaled residual sinogram $S_{res}$. To remedy these drawbacks, instead of taking the original sinogram or LI refined result as the input of $G_{sino}$, we propose a residual sinogram learning strategy for $G_{sino}$, e.g., taking $S_{res}$ as input to enhance the smoothness of preprocessed projection data, retrieving useful information from the metal trace region $M_t$ and improving the continuity at the metal trace boundary. Meanwhile, current networks contain down-sampling operations, which will cause the information loss of metal traces [36, 38]. In this work, U-Net is utilized as the backbone of $G_{sino}$. To retain sufficient information on metal traces, a mask pyramid network (MPN) [51] is introduced to explicitly feed the mask information into each layer [28, 45]. To utilize the knowledge of metal mask projection $M_p$ specifically, we concatenate $S_{res}$ and $M_p$ together as the inputs of $G_{sino}$ [45].

Thus, we have

$$M_p = \mathcal{P}(X_{metal}) \qquad (5)$$

$$M_t = \delta(M_p > 0) \qquad (6)$$

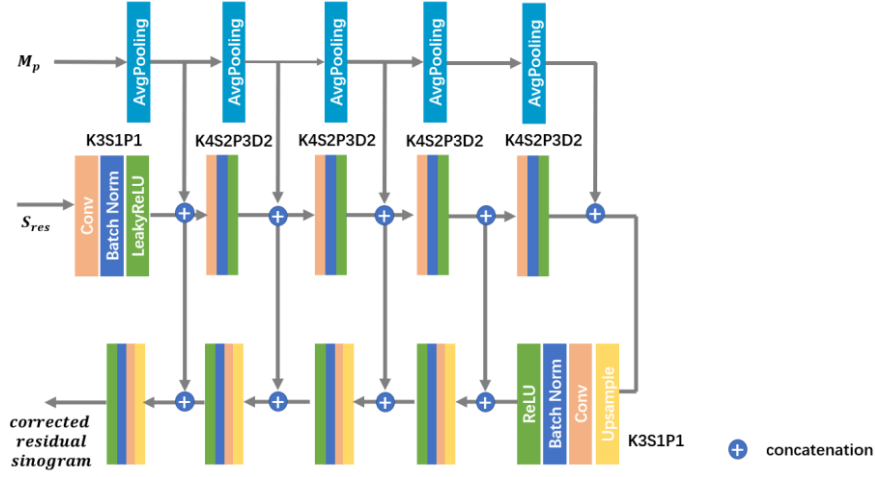

Fig. 3: An illustration of the $G_{sino}$. K: kernel, S: stride, and P: padding sizes.

where $X_{metal}$ represents the metal mask in the image domain, and $\delta(\cdot)$ is a binary indicator function.

Since our main goal is to retrieve information from the metal trace, we only refine the adaptively scaled residual sinogram in the metal trace. The details of $G_{sino}$ are shown in Fig. 3 and the corrected sinogram can be written as

$$S_{sino} = G_{sino}(S_{res}, M_p) \odot M_t + S_{LI} \qquad (7)$$

where $\odot$ stands for the elementwise multiplication. L1 loss is adopted to measure the differences between $S_{sino}$, and the ground truth $S_{gt}$ as

$$\mathcal{L}_{sino} = \| (S_{sino} - S_{gt}) \odot M_t \|_1 \qquad (8)$$

Then, $X_{sino} = \mathcal{P}^{-1}(S_{sino})$ can be obtained using an analytical reconstruction layer, which is differentiable and easily injected into neural networks.

To alleviate the secondary artifacts in the reconstructed CT image, the L1 reconstruction loss between $X_{sino}$ and the ground truth image $X_{gt}$ is utilized as

$$\mathcal{L}_{FBP} = \| (X_{sino} - X_{gt}) \odot (\mathbf{1} - M) \|_1 \qquad (9)$$

**2.2.3 Image Domain Net**

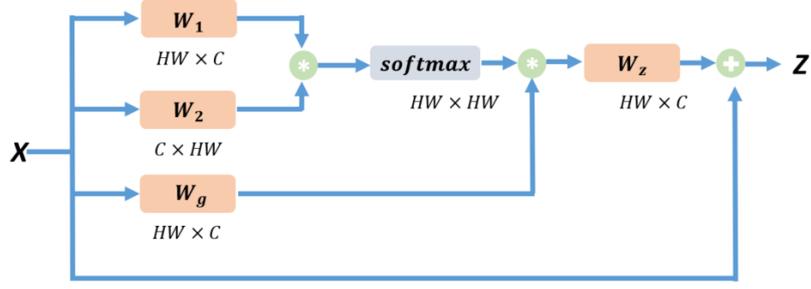

Fig. 4: An illustration of the NLN

To suppress the secondary artifacts introduced by the errors of projection data completion in $G_{sino}$, we also utilize U-Net as the backbone to enhance the reconstructed CT images. For computational efficiency, we halve the channel numbers. It is well known that convolution is a local operator whose receptive field is limited by the size of filters. Once the network is insufficiently deep, it is difficult to capture the latent features in long-range dependencies. For instance, since images have the property of self-similarity [52], and metal artifacts are nonlocal, convolution-based postprocessing methods may fail to remove the artifacts well. To tackle this problem, a nonlocal network (NLN) [53], which can capture long-range dependencies via nonlocal operations, is introduced into our proposed image domain network $G_{im}$. NLNs originate from the nonlocal means denoising method [52]. Different from nonlocal means, which performs weighted summation with similar pixels, NLN captures feature maps globally. A generic nonlocal operation is defined as

$$y_i = \frac{1}{C(x)} \sum_{j \in S} f(x_i, x_j) g(x_j) \qquad (10)$$

where $x_i$ represents the *i*-th element to be replaced, and $y_i$ is the result. $S$ represents a search window. The pairwise function $f$ computes the similarity between $x_i$ and $x_j$, which is expressed as follows:

$$f(x_i, x_j) = \exp(\theta_1(x_i)^T) \exp(\theta_2(x_j)) \qquad (11)$$

where $\theta_1(x_i) = W_1 x_i$ and $\theta_2(x_j) = W_2 x_j$ are two embeddings of feature maps, and $W_1$ and $W_2$ are the learnable weight matrices. The function $g$ serves to compute a representation of the input signal at the position of $j$. According to [53], $g$ is a linear embedding: $g(x_j) = W_g x_j$, where $W_g$ is a learned weighting matrix. $C(x)$ represents the normalization factor, which is defined as

$$C(x) = \sum_{j \in S} f(x_i, x_j) \qquad (12)$$

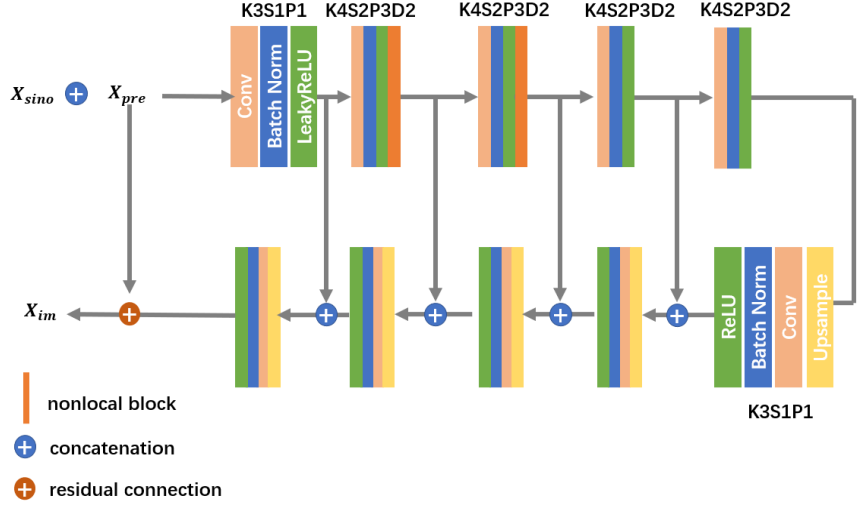

Fig. 5: An illustration of the $G_{im}$. K: kernel, S: stride, P: padding sizes and D: dilation.

To insert nonlocal operations into the neural network, a residual connection is adopted:

$$z_i = W_z y_i + x_i \qquad (13)$$

where $x_i$ denotes input data, i.e., the residual connection in ResNet [54]. In this work, the NLN module is embedded into $G_{im}$ after the second and third down-sampling steps, as depicted in Fig. 1. In NLN, $W_1$, $W_2$, $W_g$ and $W_z$ are obtained by $1 \times 1$ convolution. Fig. 4 shows the nonlocal module.

To focus on the artifact-impacted regions, $X_{sino}$ and $X_{pre}$ are concatenated as inputs of $G_{im}$. A residual learning strategy is also adopted, which is written as:

$$X_{im} = X_{pre} + G_{im}(X_{sino}, X_{pre}) \qquad (14)$$

The details of $G_{im}$ are shown in Fig. 5. $G_{im}$ is also optimized with L1 loss in the image domain:

$$\mathcal{L}_{im} = \parallel (X_{im} - X_{gt}) \odot (1 - M) \parallel_1 \qquad (15)$$

In summary, the total objective function is:

$$\mathcal{L} = \mathcal{L}_{sino} + \alpha * \mathcal{L}_{FBP} + \beta * \mathcal{L}_{im} \tag{16}$$

where $\alpha$ and $\beta$ are the weighting parameters of different components. In our experiments, we empirically set $\alpha = \beta = 1$.

3. **Experiments**

In this section, the data generation, details of neural networks, training strategies and experimental results will be shown in detail.

**3.1 Dataset**

For data simulation, we followed the procedure of [44] and used the DeepLesion dataset [55], which has high diversity and good quality. For metal mask simulation, the shape, size and positions of masks should be delicately designed to cover real clinical scenes. In this work, we employed the masks generated from [41], containing 100 manually segmented metal implants with all kinds of metal implants, such as dental fillings, spine fixed crews, hip prostheses, coiling and wires. Specifically, we randomly selected 1000 CT images from the DeepLesion dataset and 90 metal masks to synthesize 90,000 combinations in the training set. The remaining 200 CT images and 10 masks were adopted for evaluation. The original CT images were resized to $256 \times 256$ for computational efficiency. To simulate Poisson noise, a polychromatic X-ray source was employed, and the incident beam X-ray was set to $2 \times 10^7$ photons [56]. The partial volume effects and scatter were also taken into consideration. Without loss of generality, our experiments were restricted to 2D parallel-beam geometry, i.e., the sinograms of CT images were obtained by the *radon* function with MATLAB R2017b. For the sampling condition, 367 detector bins and 361 sampling views uniformly distributed from 0° to 180° were assumed. Therefore, the sinogram had a size of $367 \times 361$. Unlike [44], we truncated the CT values to [0, 4095], which better conforms to the real situation.

### 3.2 Implementation details

We trained our network in an end-to-end manner, and the model was implemented with the *PyTorch* framework [57]. The back-projection was implemented by the *numba* library in Python, which can improve the computational efficiency, aided by CUDA. The network was optimized by the Adam optimizer with the parameters $(\beta_1, \beta_2) = (0.5, 0.999)$. The learning rate was initialized to 0.0002 and halved every 20 epochs. The network was trained with 200 epochs on an NVIDIA 1080Ti GPU with 11 GB memory, and the batch size was 4.

### 3.3 Comparison with State-of-the-Art Methods

TABLE I Quantitative comparison of different methods on the simulated dataset

| Methods | Uncorrected | LI | NMAR | CNNMAR | DuDoNet | ADN | DAN-Net |
|---|---|---|---|---|---|---|---|
| PSNR | 15.33 | 30.74 | 30.83 | 32.15 | 36.82 | 33.60 | **40.61** |
| SSIM | 0.6673 | 0.9224 | 0.9270 | 0.9508 | 0.9777 | 0.9275 | **0.9872** |

The proposed DAN-Net was compared with several state-of-the-art MAR methods: linear interpolation (LI) [8], NMAR [15], CNNMAR [41], DuDoNet [44] and ADN [42]. LI and NMAR are classic methods widely used in MAR. CNNMAR is a well-known application of DL in MAR that comprehensively demonstrates the effectiveness and potential of CNN-based methods. DuDoNet is a supervised dual-domain framework in MAR that incorporates an extra sinogram enhancement network to ease the learning of the image domain. ADN is a state-of-the-art unsupervised framework in MAR that disentangles CT images corrupted by metal artifacts into an artifact-free domain and a pure artifact domain and decodes disentangled representations of artifact-free domains to artifact-suppressed images. For the LI, NMAR, CNNMAR, and ADN methods, we used publicly released codes. Because there are no public implementations of the DuDoNet method, we reimplemented it following [44].

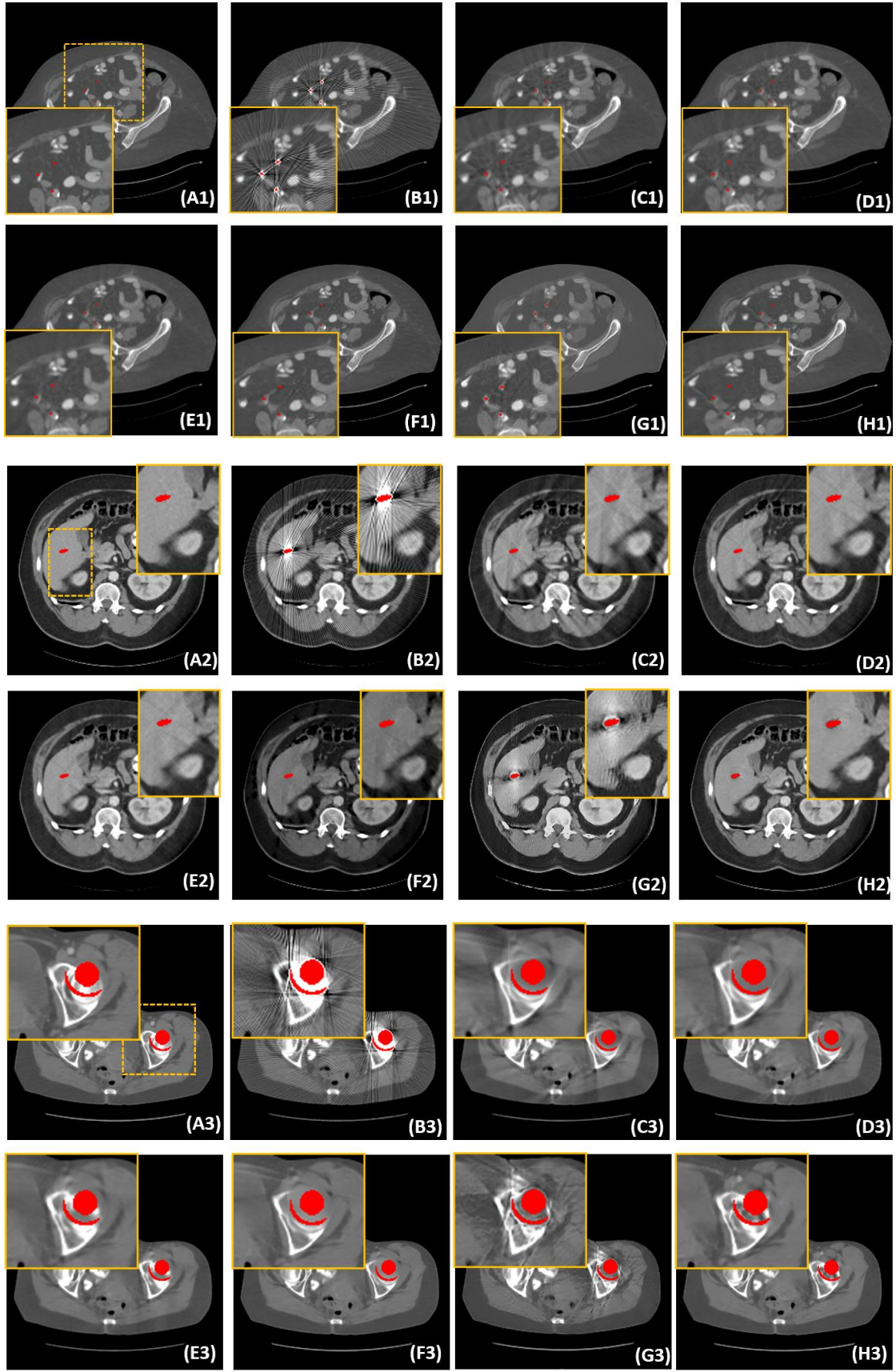

Fig. 6: Visual comparison using different methods on the simulated dataset with different metal sizes. (A1)-(A3) Reference images

with different metal sizes; (B1)-(B3) metal corrupted images; (C1)-(C3) corresponding results of LI; (D1)-(D3) corresponding results of NMAR; (E1)-(E3) corresponding results of CNNMAR; (F1)-(F3) corresponding results of DuDoNet; (G1)-(G3): corresponding results of ADN; (H1)-(H3) corresponding results of our method.

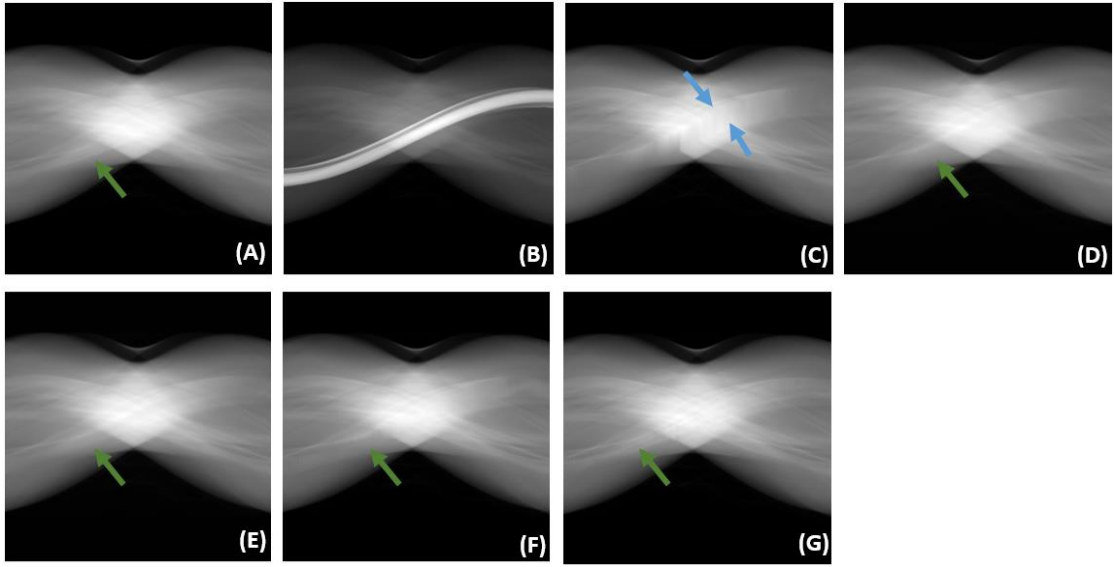

Fig. 7: Sinogram visual comparison of case 3 in Fig. 4 using different sinogram enhancement methods on the simulated dataset. (A) Reference images; (B) metal corrupted images; (C) corresponding results for LI; (D) corresponding results of NMAR; (E) corresponding results for CNNMAR; (F) corresponding results for DuDoNet; and (G) corresponding results for DAN-Net.

Structural similarity (SSIM) and peak signal-to-noise ratio (PSNR) are adopted as quantitative metrics. TABLE I lists the quantitative results obtained by calculating the mean values of both metrics on all of the test images using different methods. It is observed that the traditional MAR methods LI and NAMR significantly improve both SSIM and PSNR values compared with uncorrected CT images, and NMAR achieves better scores than LI. NMAR outperforms LI since it takes advantage of both prior images and the LI method. CNNMAR takes the outputs of different MAR methods as inputs, which is an open framework to fuse the merits of different MAR methods based on DL technology and outperforms conventional methods. ADN is an advanced unsupervised DL-based method that achieves similar performance to CNNMAR without the need for paired training data. DuDoNet and our method attain remarkable improvements on both SSIM and PSNR since they simultaneously leverage the advantages of the sinogram domain and image domain. Compared with DuDoNet, DAN-Net further raises the scores, which demonstrates the performance of our proposed method quantitatively.

For qualitative comparisons, the visual results are shown in Fig. 6, presenting three representative metallic implants with different sizes. In Fig. 6, metal-free images, metal-corrupted images and the results using different MAR methods are included. For better visualization, the simulated metal masks are colored in red. In Fig. 6, it can be seen that in the case of small metallic implants, the traditional methods, LI and NAMR, still contain some radial artifacts, while DL-based methods perform better. When metal objects get larger, LI and NAMR perform even worse. LI and NMAR introduce obvious new artifacts in Fig. 6 (C1-C3&D1-D3). Although CNNMAR suppresses secondary artifacts, distorted structures and missing tissue details can be observed in Fig. 6 (E3). Another point that needs to be mentioned is that it can be noticed in the third case that other methods fail to preserve the details around metallic implants, while DAN-Net maintains these structural details more completely. Fig. 7 shows the corresponding intermediate sinogram enhancement results. Considering that ADN is an image postprocessing method, its sinogram enhancement is not presented. In the regions indicated by the blue arrows in Fig. 7 (C), there are obvious artificial boundaries, whereas in the results of other methods, these boundaries are inconspicuous. In Fig. 7 (D-G), as indicated by the green arrows, NMAR, CNNMAR and DuDoNet generate visible differences from the reference sinogram (Fig. 7 (A)), and DAN-Net achieves the most visibly consistent sinogram with the reference.

**3.4 Clinical study**

To verify the performance of proposed DAN-Net in a clinical scenario, two clinical CT images with small and large metal artifacts were tested. In this experiment, the metal artifacts were empirically segmented using 2000 HU as the threshold. The test images were normalized to the same range as the training data.

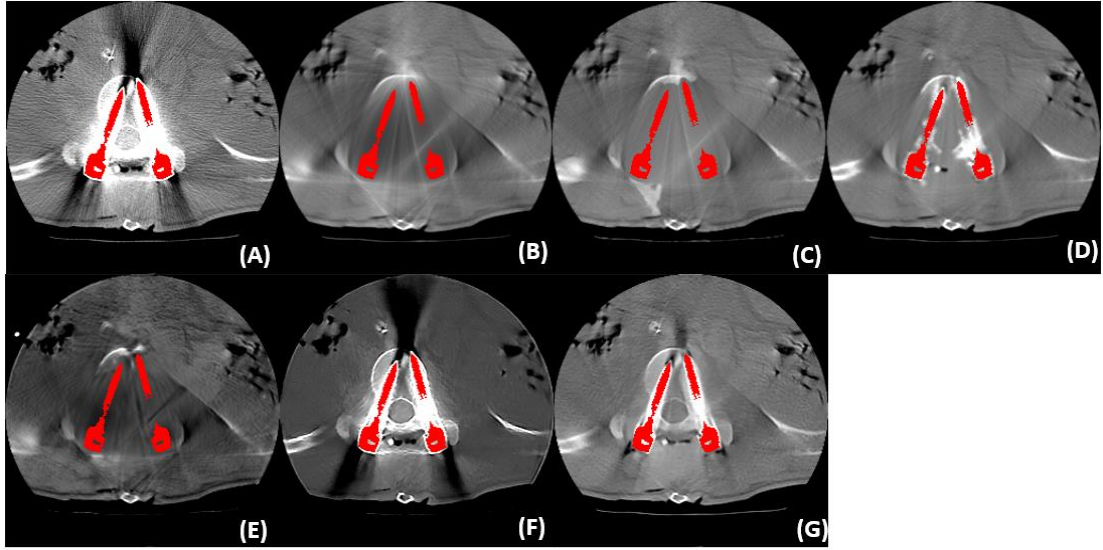

Fig. 8: Visual comparison with different MAR methods on a clinical CT image. A1-G1 and A2-G2 represent uncorrected CT images and corrected results using LI, NMAR, CNNMAR, DuDoNet, ADN and DAN-Net.

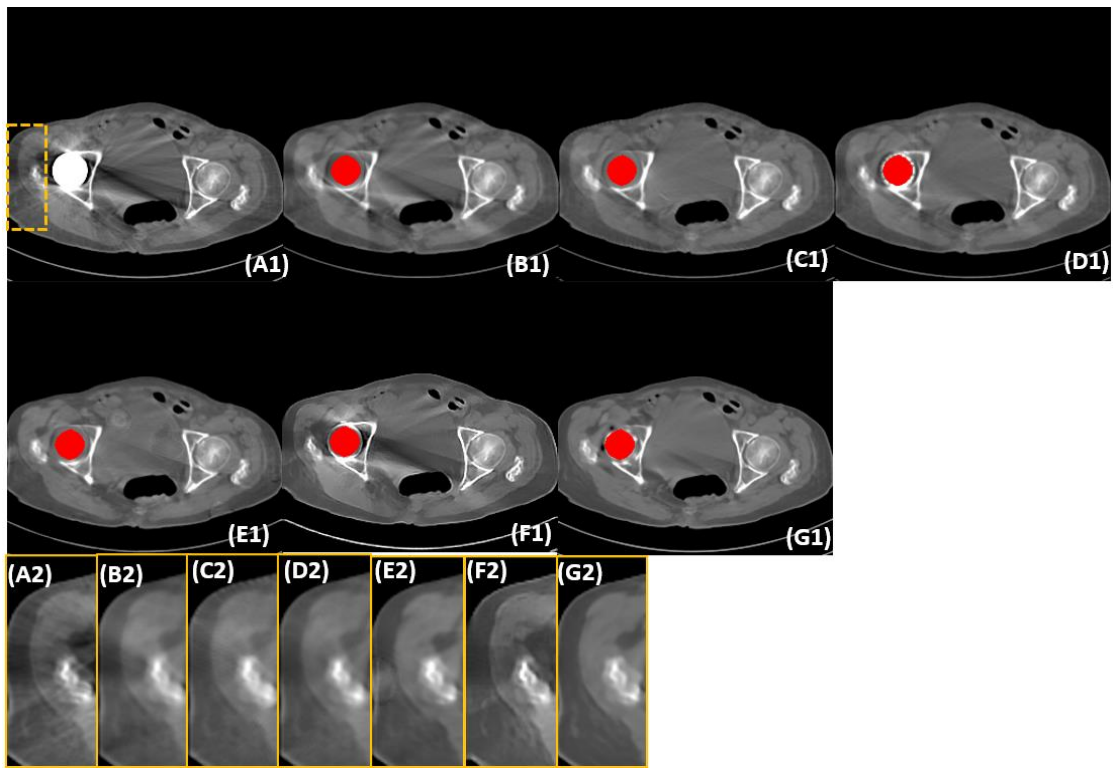

Fig. 9: Visual comparison with different MAR methods on a clinical CT image. A1-G1 and A2-G2 represent uncorrected CT images and corrected results using LI, NMAR, CNNMAR, DuDoNet, ADN and DAN-Net.

Fig. 8 and Fig. 9 present the MAR results using different methods. It is observed that DAN-Net suppresses most of the metal artifacts and preserves the fine-grained anatomical structures around the metals, which supplies coherent results to the simulated data and demonstrates the potential for real clinical application. Meanwhile, the performance of most MAR methods is dependent on the previous results of segmentation, and our method will also benefit from a more accurate segmentation algorithm.

TABLE II: Quantitative comparison of different variants of our method on the simulated dataset.

| Methods | Uncorrected | Sino-Net | Res-Sino-Net | IM-Net | Non-local-IM-Net | Ma-Dual-Net | DAN-Net |
|---------|-------------|----------|--------------|--------|------------------|-------------|---------|
| PSNR    | 15.33       | 31.43    | 31.71        | 33.79  | 34.75            | 34.15       | **40.61** |
| SSIM    | 0.6673      | 0.9232   | 0.9494       | 0.9520 | 0.9720           | 0.9597      | **0.9872** |

**4. Ablation Study**

In this section, we investigate the effectiveness of different modules of the proposed DAN-Net. The ablation study configurations are listed as follows:

1) Sino-Net: the sinogram-domain network without residual learning;

2) Res-Sino-Net: the sinogram-domain network with residual learning;

3) IM-Net: the image-domain network without a nonlocal module;

4) Nonlocal-IM-Net: the image-domain network with the nonlocal module;

5) Ma-Dual-Net: a dual-domain network with sinogram-domain residual learning and an image-domain nonlocal module without an adaptively scaled sinogram; and

6) DAN-Net: the same architecture as Ma-Dual-Net with an adaptively scaled sinogram.

The quantitative results of the ablation study are given in TABLE II and the visual results are shown in Fig. 10 and Fig. 11.

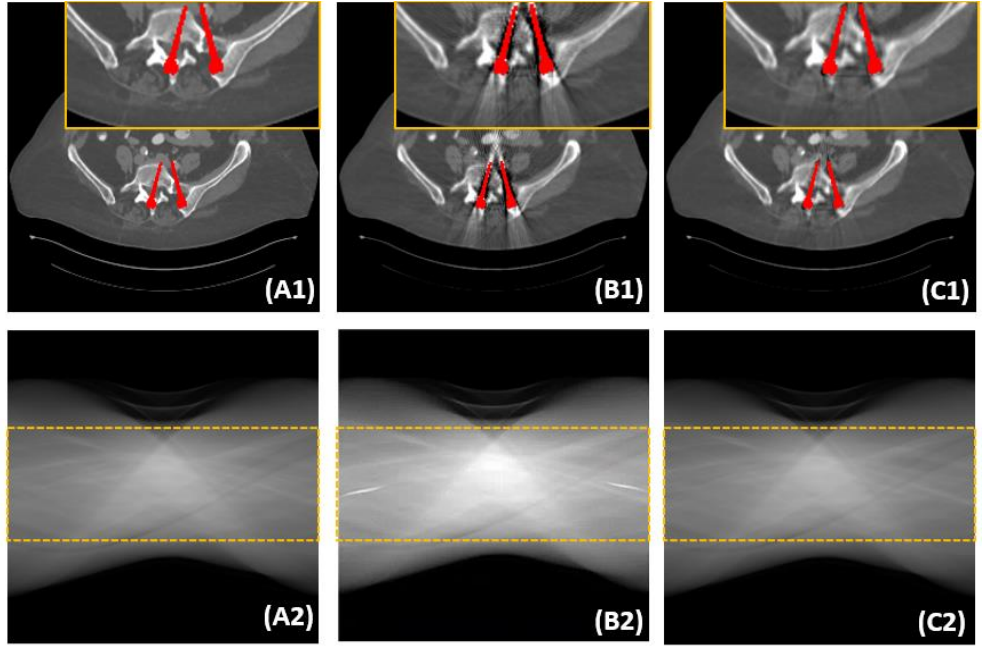

Fig. 10: Sinograms and corresponding reconstructions with sinogram-domain enhancement methods. The simulated metal masks are colored red. (A1) and (A2): ground truth. (B1) and (B2): Sino-Net. (C1) and (C2): Res-Sino-Net.

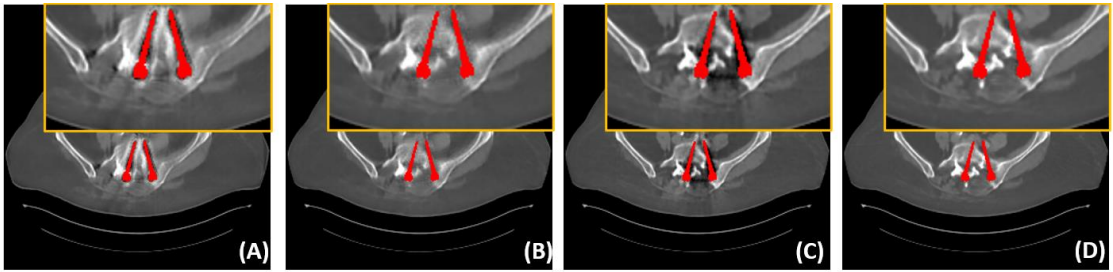

Fig. 11: Reconstructions using image-domain and dual-domain enhancement methods. The simulated metal masks are colored in red. The reference image is Fig. 8 (A1). (A): IM-Net, (B): Nonlocal-IM-Net, (C): Ma-Dual-Net and (D): DAN-Net.

**4.1 Effect of sinogram-domain residual learning.**

To evaluate the performance of our residual sinogram learning strategy, a neural network that takes the original sinogram as input to complete the projection data within the metal trace was trained. In Table II, it is obvious that the residual sinogram learning strategy significantly improves the SSIM and PSNR values. The visual results are shown in Fig. 10 (C) and (D). There are evident artifacts in the result of Sino-Net, experimentally demonstrating that it is difficult for a convolutional neural network to transform two different data distributions to the same distribution in the sinogram domain. In contrast, in the Res-

Sino-Net results, residual information is recovered from adaptively scaled projection data within the metal trace, thereby easing network learning.

**4.2 Effect of image-domain nonlocal module.**

To further suppress artifacts in the image domain, a nonlocal U-Net architecture is adopted. To validate the effectiveness of this modification, a neural network is trained to refine the CT images directly without the nonlocal module. The qualitative comparison is shown in TABLE II, from which we can see that the nonlocal-IM-Net has higher SSIM and PSNR values than IM-Net. For the qualitative comparison, it is observed that artifacts are better suppressed in the results of Nonlocal-IM-Net than IM-Net in Fig. 11 (A) and (B).

**4.3 Effect of preprocessing.**

In this subsection, the impact of adaptive scaling is sensed. Ma-Dual-Net takes the original sinogram and corresponding reconstructed image as the inputs, and ours takes the adaptively scaled sinogram and corresponding reconstructed image as the inputs. In Table II, our approach outperforms Ma-Dual-Net in quantitative aspects. The visual comparison is also presented in Fig. 11 (C) and (D), in which we can observe that our method retrieves many more structural details around the metallic implants.

**5. Discussions and conclusion**

Due to the insertion of metals, the imaging quality of CT images will significantly degrade. Over the past few decades, a large number of MAR methods have been proposed to alleviate the effects of metal artifacts in CT images. In conventional methods, projection data in the metal trace are regarded as missing, and MAR is formulated as an image interpolation or inpainting problem. Some interpolation methods, such as linear interpolation and cubic polynomial interpolation, are applied to fill the missed projection data. Nonetheless, since most interpolation methods cannot guarantee continuity near the interpolation boundary, there are apparent borderlines in the corrected sinogram, and secondary artifacts appear. Furthermore, since the projection data in the metal trace are simply abandoned and replaced with the value estimated with the data outside the metal trace, the information within the metal trace is lost, leading to the loss of tissue details around the metal in the reconstructed CT image. Therefore, not only secondary artifacts but also details are lost in interpolation-based methods. In practice, it is difficult for single-domain methods to achieve both goals simultaneously [44]. However, interpolation-based methods can generate a proper initial estimation for DL-based methods, which has been employed in several works. In our work, we also introduce this technique.

In this work, we combine the advantages of conventional MAR approaches and DL-based methods to further improve the performance. Although we adopt the same end-to-end training strategy, there are some significant differences. To restrain artifacts and maintain tissue details more efficiently, adaptive scaling on the original projection data in the metal trace is applied. Then, the preprocessed sinogram and corresponding reconstructed CT images are utilized as the inputs of our network. Because metal has a much higher attenuation coefficient, the projection data inside and outside the metal trace can be regarded as two different data distributions. It is difficult to convert two different data distributions to a unified distribution for normal networks. To tackle this problem, a residual learning strategy that only modifies the metal trace region values of the adaptively scaled sinogram is used. To alleviate the new artifacts introduced in image domain enhancement, we propose a novel nonlocal U-Net architecture that can capture long-range dependencies to suppress metal artifacts.

However, there are some limitations to our work, and we will dedicate ourselves to solving them in the future. In an end-to-end training manner, it is preferable to obtain the adaptive parameter by learning instead of through a manual setting. Fortunately, the subsequent filtering can reduce the influence of inaccurate parameter selection according to [50]. In the future, we will investigate how to integrate this parameter learning into the model to minimize human interference.

We trained and evaluated our networks on simulated datasets, and few clinical CT images were used to validate the effectiveness of our model. In the future, we will collect large-scale clinical images to evaluate the performance of our method in the clinical scenario more comprehensively and systematically.

**Declaration of Competing Interest**

The authors declare that they have no known competing financial interests or personal relationships that could have appeared to influence the work reported in this paper.

**CRediT authorship contribution statement**

**Tao Wang:** Conceptualization, Methodology, Writing - original draft. Investigation, Software. **Wenjun Xia:** Writing - review & editing, Formal analysis. **Yongqiang Huang:** Writing - review & editing. **Huaiqiang Sun:** Data curation. **Yan Liu:** Writing - review & editing, Funding acquisition. **Hu Chen:** Writing - review & editing, Funding acquisition. **Jiliu Zhou:** Data curation, Funding acquisition, Project administration. **Yi Zhang:** Methodology, Writing - review & editing, Funding acquisition, Project administration.

**Acknowledgements**

This work was supported in part by the National Natural Science Foundation of China under Grant 61871277, 61902264 and in part by the Sichuan Science and Technology Program under Grant 2021JDJQ0024, 2019YFS0125.